\documentclass[dvipsnames]{aa}
\usepackage[varg]{txfonts}
\usepackage{natbib}
\bibpunct{(}{)}{;}{a}{}{,} 
\usepackage{hyperref}
\usepackage{enumitem}
\usepackage{amsmath}
\usepackage{etoolbox}
\usepackage{tikz}
\newrobustcmd*{\mysquare}[1]{\tikz{\filldraw[draw=#1,fill=#1] (0,0) rectangle (0.18cm,0.18cm);}}
\newrobustcmd*{\mycircle}[1]{\tikz{\draw[draw=#1] (0,0) circle [radius=0.1cm];}}
\newrobustcmd*{\mybullet}[1]{\tikz{\filldraw[draw=#1,fill=#1] (0,0) circle [radius=0.1cm];}}
\DeclareMathAlphabet{\mathitbf}{OML}{cmm}{b}{it}
\newcommand{\Avec}{\mathitbf{A}}

\newcommand{\Apvec}{\mathitbf{A}_{\mathrm{0}}}
\newcommand{\Bvec}{\mathitbf{B}}
\newcommand{\Bx}{\mathit{B_x}}
\newcommand{\By}{\mathit{B_y}}
\newcommand{\Bz}{\mathit{B_z}}

\newcommand{\Bhor}{\mathit{B_{\rm h}}}

\newcommand{\Bpvec}{\mathitbf{B}_{\mathrm{0}}}
\newcommand{\Bps}{\mathitbf{B}_{\mathrm{0},s}}
\newcommand{\Bpns}{\mathitbf{B}_{\mathrm{0},ns}}
\newcommand{\Bjvec}{\mathitbf{B}_J}
\newcommand{\Bjs}{\mathitbf{B}_{\mathrm{J},s}}
\newcommand{\Bjns}{\mathitbf{B}_{\mathrm{J},ns}}

\newcommand{\Binput}{\mathitbf{B}_{\mathrm{in}}}

\newcommand{\Wvec}{\mathitbf{W}}
\newcommand{\WvecE}{\mathitbf{W}_{\rm EMP}}
\newcommand{\WvecH}{\mathitbf{W}_{\rm HMI}}

\newcommand{\Etot}{E}
\newcommand{\Epot}{E_{\mathrm{0}}}

\newcommand{\Ediv}{E_{\mathrm{div}}}
\newcommand{\Ejs}{E_{\mathrm{J,s}}}
\newcommand{\Ejsprime}{\Ejs/\Etot}
\newcommand{\Eps}{E_{\mathrm{0,s}}}

\newcommand{\Ejns}{E_{\mathrm{J,ns}}}
\newcommand{\Epns}{E_{\mathrm{0,ns}}}
\newcommand{\Emix}{E_{\mathrm{mix}}}
\newcommand{\Edivprime}{E_{\rm div}/E}
\newcommand{\Emixprime}{|E_{\rm mix}|/\Ejs}
\newcommand{\efo}{\epsilon_{\rm force}}
\newcommand{\thetaj}{\theta_J}
\newcommand{\wa}{w_f}
\newcommand{\wb}{w_d}

\newcommand{\wperp}{w_\perp}
\newcommand{\wpara}{w_\parallel}
\newcommand{\fiavg}{\langle|f_i|\rangle}

\newcommand{\hpj}{H_{\mathrm{PJ}}}
\newcommand{\hj}{H_{\mathrm{J}}}
\newcommand{\hjprime}{|\hj|/|\hv|}
\newcommand{\hv}{H_{\mathcal{V}}}
\newcommand{\mhjprime}{\langle|\hj|/|\hv|\rangle}
\newcommand{\degree}{^\circ}
\newcommand{\ie}{{\it i.e.}}
\newcommand{\eg}{{\it e.g.}}
\title{Deducing the reliability of relative helicities  from nonlinear force-free coronal models}
\author{Julia K. Thalmann\inst{\ref{inst1}}\and X. Sun\inst{\ref{inst2}}\and
K. Moraitis\inst{\ref{inst3},\ref{inst4}}\and M. Gupta\inst{\ref{inst1}}}
\institute{%
University of Graz, Institute of Physics/IGAM, Universit\"atsplatz 5, 8010 Graz, Austria\label{inst1} 
\email{julia.thalmann@uni-graz.at}
\and 
Institute for Astronomy, University of Hawaii at Manoa, Pukalani, HI 96768-8288, USA\label{inst2}
\and
LESIA, Observatoire de Paris, Universit{\'e} PSL, CNRS, Sorbonne Universit{\'e}, Universit{\'e} de Paris, 5 place Jules Janssen, 92195 Meudon, France\label{inst3}
\and
University of Ioannina, Department of Physics, Section of Astrogeophysics, Ioannina, Greece \label{inst4}
}
\date{Received July 15, 2020 / Accepted October 12, 2020 }

\abstract{
{\it Aims:} We study the relative helicity of active region (AR) NOAA~12673 during a ten-hour time interval centered around a preceding X2.2 flare (SOL2017-09-06T08:57) and also including an eruptive X9.3 flare that occurred three hours later (SOL2017-09-06T11:53). In particular, we aim for a reliable estimate of the normalized self-helicity of the current-carrying magnetic field, the so-called helicity ratio, $\hjprime$, a promising candidate to quantity the eruptive potential of solar ARs.\\
{\it Methods:} Using {\it Solar Dynamics Observatory} Helioseismic and Magnetic Imager vector magnetic field data as an input, we employ nonlinear force-free (NLFF) coronal magnetic field models using an optimization approach. The corresponding relative helicity, and related quantities, are computed using a finite-volume method. From multiple time series of NLFF models based on different choices of free model parameters, we are able to assess the spread of $\hjprime$, and to estimate its uncertainty.\\
{\it Results:} 
In comparison to earlier works, which identified the non-solenoidal contribution to the total magnetic energy, $\Edivprime$, as selection criterion regarding the required solenoidal quality of magnetic field models for subsequent relative helicity analysis, we propose to use in addition the non-solenoidal contribution to the free magnetic energy, $\Emixprime$. As a recipe for a reliable estimate of the relative magnetic helicity (and related quantities), we recommend to employ multiple NLFF models based on different combinations of free model parameters, to retain only those that exhibit smallest values of both $\Edivprime$ and $\Emixprime$ at a certain time instant, to subsequently compute mean estimates, and to use the spread of the individually contributing values as an indication for the uncertainty.
} 


\begin{document}

\titlerunning{Reliability of relative helicities}
\maketitle

\section{Introduction}\label{sec:introduction} 

Based on the Gauss linking number, magnetic helicity is a measure of the level of entanglement of magnetic field lines within a magnetized plasma \citep{1969JFM....35..117M}. It is strictly conserved within the ideal magneto-hydrodynamic paradigm \citep{1958PNAS...44..833W}, and its dissipation  is relatively weak even in non-ideal magneto-hydrodynamics \citep{1974PhRvL..33.1139T}, the latter even in the presence of strong non-ideal effects \citep{2015A&A...580A.128P}. In the context of solar eruptivity, this favorable property  can explain the existence of plasma ejecta in order to prevent infinite accumulation within the solar atmosphere \citep{1994GeoRL..21..241R,1996SoPh..167..217L}. 

The basic formulation of magnetic helicity lacks gauge transform invariance for magnetically open systems, \ie, it is not directly applicable to studies of the solar corona since magnetic flux is continuously penetrating the coronal volume through the solar surface. To circumvent this limitation, \cite{1984JFM...147..133B} and \cite{1984CPPCF...9..111F} defined the so-called relative magnetic helicity as
\begin{equation}
        \hv=\int_\mathcal{V}\left(\Avec+\Apvec\right)\cdot\left(\Bvec-\Bpvec\right) {\rm ~d}\mathcal{V}, \label{eq:hv}
\end{equation}
a gauge-invariant quantity related to the magnetic helicity within a volume, $\mathcal V$, bounded by a surface, $\partial\mathcal V$. Here $\Bvec$ and $\Bpvec$ are the 3D magnetic field under study and a reference field, respectively, while $\Avec$ and $\Apvec$ are the vector potentials satisfying $\Bvec=\nabla\times\Avec$ and $\Bpvec=\nabla\times\Apvec$, respectively.

As its name implies, the relative helicity  expresses the helicity of a magnetic field with respect to a reference field, $\Bpvec$,  which shares the normal component of the studied field $\Bvec$ on $\partial\mathcal{V}$. Generally, $\Bpvec$ is chosen to be a potential (current-free) field \citep[see][for an alternative choice]{2014ApJ...787..100P}. For practical cases, \cite{2012SoPh..278..347V} demonstrated the validity and physical meaningfulness to compute (and track in time) $\hv$ by evaluating \href{eq:hv}{Eq.~(\ref{eq:hv})} in order to characterize (the evolution of) a magnetic system.

 \cite{1999PPCF...41B.167B} decomposed $\hv$ into two separately gauge-invariant quantities
\begin{eqnarray}
        \hj&=&\int_\mathcal{V}\left(\Avec-\Apvec\right)\cdot\left(\Bvec-\Bpvec\right) {\rm ~d}\mathcal{V}, \label{eq:hj}\\
        \hpj&=&2\int_\mathcal{V}\Apvec\cdot\left(\Bvec-\Bpvec\right) {\rm ~d}\mathcal{V}, \label{eq:hpj}
\end{eqnarray}
so that $\hv=\hj+\hpj$. Here $\hj$ is the magnetic helicity in the volume associated with the electric current, and $\hpj$ is the helicity associated with the component of the field that is threading $\partial\mathcal V$. Since $\Bpvec$ is designed to share the normal component with $\Bvec$ on $\partial\mathcal{V}$, consequently $\hv$,  $\hj$, and $\hpj$ are all independently gauge invariant. For a pilot study of the time evolution of these quantities in the solar context see \cite{2014SoPh..289.4453M}. For completeness we note that, unlike $\hv$ in \href{eq:hv}{Eq.~(\ref{eq:hv})}, $\hj$ and $\hpj$ are not conserved quantities as a gauge-invariant transfer term between them dominates their dynamics \citep{2018ApJ...865...52L}.

In particluar,  $\hj$ in \href{eq:hj}{Eq.~(\ref{eq:hj})} attracts attention as it provides additional information compared to $\hv$. More precisely, the so-called helicity ratio, $\hjprime$, appeared as a promising candidate in characterizing the eruptive potential of the underlying magnetic structure. This was noted not only from numerical simulations \citep[\eg,][]{2017A&A...601A.125P,2018ApJ...863...41Z,2018ApJ...865...52L}, but also from application to solar observations \citep[][]{2018ApJ...855L..16J,2019A&A...628A..50M,2019ApJ...887...64T}.

Magnetic helicity studies of solar observations are often performed based on nonlinear force-free (NLFF) coronal magnetic field extrapolations, \ie, the
numerical solution of 
\begin{eqnarray}
        \left(\nabla\times\Bvec\right)\times\Bvec &=& \mathbf{0} \label{eq:ff1},\\
        \nabla\cdot\Bvec&=&0,\label{eq:ff2}
\end{eqnarray}
where $\Bvec$ represents the 3D coronal magnetic field, subject to the measured surface magnetic field as a boundary condition \citep[for a review see, \eg,][]{2012LRSP....9....5W}. For instance, \cite{2018ApJ...855L..16J} used a magneto-frictional method to solve \href{eq:ff1}{Eqs.~(\ref{eq:ff1})} and \href{eq:ff2}{(\ref{eq:ff2})} , while the works of \cite{2019A&A...628A..50M} and \cite{2019ApJ...887...64T} were based on an optimization approach. Whatever method used, unavoidable numerical errors prevent the exact fulfillment of \href{eq:ff1}{Eqs.~(\ref{eq:ff1})} and \href{eq:ff2}{(\ref{eq:ff2})}. In particular, the level of solenoidality of the obtained NLFF solution is highly important for relative helicity computations (\citealt{2013A&A...553A..38V}; and see \href{ss:fv_helicity}{Sect.~\ref{ss:fv_helicity}} for details).

Recently, \cite{2019A&A...628A..50M} (hereafter M19) studied the eruptivity of NOAA~12673 which produced the two strongest flares of Solar Cycle 24 on 6~September~2017. A confined X2.2 flare started at 08:57~UT (SOL2017-09-06T08:57), and an eruptive X9.3 flare followed three hours later (start time 11:53~UT; SOL2017-09-06T11:53). In their study a ten-hour time interval centered around the X2.2 flare of 6~September that also includes the X9.3 flare was used. Their analysis used a mix of NLFF solutions, based on the optimization method of \cite{2012SoPh..281...37W}, a specialization of the method originally described in \cite{2010A&A...516A.107W} for the application to {\it Solar Dynamics Observatory}  Helioseismic and Magnetic Imager data (hereafter W12), and on its original predecessor \citep[][hereafter W04]{2004SoPh..219...87W}, using standard (free) model parameter settings. They argued that  employing NLFF models based on different code versions   optimizes the final NLFF time series used to compute the coronal relative helicity, when retaining only those that perform best in terms of solenoidality. Thus, at each time instant within the studied time series, they checked the solenoidal quality of the W04 and W12 solutions, and used the particular NLFF solution of highest solenoidal quality, \ie, that with the smallest value of $\nabla\cdot\Bvec$. Time instances where none of the employed models were providing an acceptably small level of solenoidality were discarded entirely. The time evolution of $\hjprime$ that resulted based on this pre-selection of NLFF models depicts an increase in $\hjprime$ to values $>0.15$ prior to the X-class flares, as well as corresponding decreases in the course of the flares (see their Fig.~7).

\cite{2015ApJ...811..107D} delivered, although as a secondary result, the first comparative analysis of relative helicity computations based on different NLFF methods, picturing consistent relative helicity estimates as a challenging yet achievable task. In that work the W12-based NLFF solutions were found to deliver distinctly different values for the relative helicity in comparison to those deduced from other NLFF methods (a magneto-frictional and three Grad-Rubin methods), which was explained by the insufficient solenoidal quality of the NLFF model. This issue was found  to be linked to the use of standard choices of (free) model parameters, as suggested in W12, which resulted in NLFF solutions with non-solenoidal errors on the order of the inherent free magnetic energy (see their Fig.~7b). It was also shown, however, that an alternative W12-based NLFF solution based on an adjusted set of model parameters resulted in a significant improvement of the solenoidal quality, and hence the corresponding relative helicity computation (see Appendix of their work). 

The above works suggests that there is great potential for improving the accuracy of relative helicity estimates, based on different model parameter choices and/or particular versions of the optimization approach. Since it has been shown that the W12 method delivers NLFF solutions with a higher degree of force-freeness and lower solenoidal level in comparison to the W04 method \citep[see Table~2 in][]{2012SoPh..281...37W}, we restrict ourselves to making use of the W12 method and employ a number of NLFF solutions based on different choices for the adjustable (free) model parameters. In order to make our results comparable to the previous study of M19, we use the same vector magnetic field data as input to NLFF modeling. We attempt to assess the resulting spread of the relative helicity and related quantities, most importantly that of $\hjprime$, for this particular AR and time interval in relation to the particular NLFF model parameters used. On that basis, we aim to provide a recipe for a realistic computation of the relative helicity, including an appropriate estimation of related uncertainties.

\section{Data and methods}

\subsection{Vector magnetic field data}\label{ss:data}

In our study we use the data set originally designed to study the eruptivity of NOAA AR 12673 in M19, originally based on the 12-min cadence {\sc hmi.sharp\_720s} data product, constructed from polarization measurements of the {\it Solar Dynamics Observatory}  Helioseismic and Magnetic Imager \citep[{\it SDO}/HMI;][]{2012SoPh..275..207S}. A field of view covering $320\times320$ pixels was extracted from the full-disk {\sc hmi.sharp\_720s} data, centered at the Carrington coordinates ($118.4\degree$,$-9.2\degree$). A cylindrical equal area (CEA) projection was applied to the chosen subfield of the {\sc hmi.sharp\_720s} data vector field data, following the description in \cite{2013arXiv1309.2392S}. The resulting CEA-remapped field vector ($B_r$,$B_\theta$,$B_\phi$) was binned by a factor of two to a resolution of 0.06~degree ($\sim$720~km at disk center). In this way a total of 17 CEA vector magnetic field maps were constructed, covering the time span 2017-Sep-06 04:00~UT -- 13:00~UT, around two major flares hosted by NOAA~12673 (a confined X2.2 flare that peaked at 09:10~UT, and an eruptive X9.3 flare that peaked at 12:02~UT). 

\subsection{NLFF modeling}\label{ss:nlff}

Based on the vector magnetic field data described in \href{ss:data}{Sect.~\ref{ss:data}}, we compute a series of NLFF equilibria for each of the 17 time instances. NLFF modeling involves one computational task at least (optimization; see \href{sss:optim}{Sect.~\ref{sss:optim}}), and two computational tasks at most (a preprocessing step, see \href{sss:prepro}{Sect.~\ref{sss:prepro}}, followed by subsequent optimization).  Preprocessing is necessary, because the vector magnetic field data deduced from polarization measurements at photospheric levels does not conform with the force-free criteria \citep{1989SoPh..120...19A}. 

\subsubsection{Preprocessing}\label{sss:prepro}

During preprocessing, the measured 2D vector magnetic field data is modified in order to obtain a vector field that is (more) force-free consistent. The preprocessing method of \cite{2006SoPh..233..215W} minimizes a function of the form
\begin{eqnarray}
        L_{\rm pp} &=& \mu_1 L_{\rm pp,1} + \mu_2 L_{\rm pp,2} + \mu_3 L_{\rm pp,3} + \mu_4 L_{\rm pp,4} , \label{eq:l_pp}
\end{eqnarray}
where the individual contributions $L_{\rm pp,i}$ are summed over all grid points of the 2D photospheric grid, and are weighted individually by the corresponding pre-factors $\mu_i$. In discretized form, $L_{\rm pp,1} $ is the square of the total magnetic force, and $L_{\rm pp,2} $ the square of the total magnetic torque; $L_{\rm pp,3} $ measures the difference between the preprocessed and original input field, and $L_{\rm pp,4} $ reduces small-scale variations in the measured field (applying a Laplacian smoothing). 

In the solar context, preprocessing aims to approximate the chromospheric magnetic field, assumed to be more force-free consistent than the photospheric magnetic field. In  \cite{2008SoPh..247..249W} a realistic model active-region is used to test the effect of preprocessing. In addition to the original scope of that study, the preprocessing has been shown to remove non-magnetic forces in the model photosphere and to yield a chromospheric-like model field (see their Table~1 and Fig.~2). In particular, the smoothing term, $L_{\rm pp,4}$,  in \href{eq:l_pp}{Eq.~(\ref{eq:l_pp})} is physically motivated by the desire to approximate the characteristic spatial scales at a chromospheric level, \ie, to remove all scales below super-granular diameter \citep{2006SoPh..233..215W}. The smoothing term  naturally competes with the changes to the data due to the terms assigned to enforce force-free compatibility ($L_{\rm pp,1}$ and $L_{\rm pp,2}$). Since $L_{\rm pp,1}$ and $L_{\rm pp,2}$ are weighted distinctly stronger (usually, $\mu_1=\mu_2=1$ and $\mu_4\ll\mu_1$), $L_{\rm pp,4}$ may have only a limited effect. As a consequence, small spatial scales might actually be enhanced \citep[see corresponding remarks in Sect.~8.1 of ][]{2013A&A...553A..38V}.

The application of preprocessing prior to optimization has the desired effect of   delivering NLFF solutions of higher quality, independent of the particular NLFF method used, but appeared advantageous especially for NLFF methods that rely on numerical differentiation \citep[see Sect.~7.3 of][]{2008SoPh..247..269M}. It improves the final NLFF solution, both in terms of force- and divergence-freeness, naturally because of the more force-free consistent nature of the preprocessed data (see Table~2 in W12 and also Table~2 in \citealt{2008SoPh..247..249W}).

The recommended (standard) relative weightings suggested in W12 are $(\mu_1,\mu_2,\mu_3,\mu_4)=(1,1,10^{-3},10^{-2})$, with the weightings $\mu_1$ and $\mu_2$ set several orders of magnitude larger than $\mu_3$ and $\mu_4$. This is because meeting the (nearly) vanishing total magnetic force and torque is essential, while the nearness to the actually observed data and its smoothness are desired (but secondary) requirements. In our work, we use the suggested setting $\mu_1=\mu_2=1$, and inspect separately the effect of enforcing nearness to the observed data ($\mu_3=10^{-3}$ versus\ $\mu_3=0$) and that of smoothing ($\mu_4=10^{-2}$ versus\ $\mu_4=0$). Though the advantageous effect of smoothing onto the quality of the resulting NLFF solution is known, its impact on relative helicity computation is still unclear. Using the just presented extreme choices of corresponding relative weightings, we are able to clarify the relative influences of the actually observed data and smoothing.

\subsubsection{Optimization}\label{sss:optim}

In order to perform the NLFF optimization, we apply the method of W12, \ie, we combine the improved optimization scheme of \cite{2010A&A...516A.107W} and a multiscale approach \citep[][]{2008JGRA..113.3S02W}. In our work we apply a three-level multiscale approach to the (non-)preprocessed vector magnetic field data. The optimization approach is designed such that the function 
\begin{eqnarray}
        L &=& \int_V \left( \wa\,\frac{|\left(\nabla\times\Bvec\right)\times\Bvec|^2}{B^2}+\wb\,|\nabla\cdot\Bvec|^2\right)\,{\rm d}v \nonumber \\
        &+& \nu \int_S\left(\Bvec-\Binput\right)\cdot\Wvec\cdot\left(\Bvec-\Binput\right)\,{\rm d}s, \label{eq:l_ff}
\end{eqnarray}
is minimized yielding the volume-integrated Lorentz force and divergence to become small.

The surface term in \href{eq:l_ff}{Eq.~(\ref{eq:l_ff})} allows deviations between the NLFF solution, $\Bvec$, and the magnetic field information at the lower boundary, $\Binput$, in order to find a more force-free solution. The deviation from $\Binput$ is controlled by the diagonal error matrix (\ie, the non-diagonal elements are zero), $\Wvec$, which allows it to account for the uncertainties on each component of the magnetic field, and in each pixel, separately. Here $\Binput$ may be either  a directly measured and force-free consistent or a preprocessed vector magnetic field. The following model parameters  can be freely assigned in \href{eq:l_ff}{Eq.~(\ref{eq:l_ff})}:
\begin{itemize}[itemsep=1pt]
\item[--] Separate weightings of the volume-integrated force ($\wa$) and divergence ($\wb$). In the original notation of W12 they are set as $\wa=\wb=1$. 
\item[--] The injection speed of the lower boundary, \ie, the relative importance of the surface term in \href{eq:l_ff}{Eq.~(\ref{eq:l_ff})}, is controlled by $\nu$. W12 tested $\nu$ in the range $10^{-4}$--$10^{-1}$ and found $\nu=10^{-3}$ to represent a most qualified choice for the application to HMI data. This is because higher values yield a lower force-free and solenoidal quality of the resulting NLFF solution, while lower values yield little corresponding improvement, despite drastically increased computation times. Therefore, we use a value of $\nu=10^{-3}$  in our study, as has also been used as in the work of M19.
\item[--] The components $\wpara$ (controlling the weighting of the horizontal field components $\Bx$ and $\By$) and $\wperp$ (controlling the weighting of the vertical field component $\Bz$) of the diagonal error matrix, $\Wvec$, can be defined in different ways; in the most sophisticated of which each individual pixel may be weighted based on the actual HMI measurement uncertainties. Only recently such an attempt has been presented in M19, who chose 
\begin{eqnarray}
        \wpara = \wperp = 
        \left\{
        \begin{array}{ll}
        0.01 + 0.99 \exp \left(-\frac{\sigma_B}{0.03 B}\right), & \text{if } B\ge200{\rm ~G} \\
        0.01, & \text{if } B<200{\rm ~G} \\
        \end{array}
        \right.,\label{eq:WvecH}
\end{eqnarray}
where $B$ denotes the magnetic field strength and $\sigma_B$ is the total magnetic field variance from inversion fitting. The authors thus assumed a typical noise threshold of 200~G and a typical value of 0.03 for $\sigma_B/B$. This particular weighting was designed in order to compensate for low-quality inversion results, covering regions of strong magnetic field and spreading out in the later frames of the time series. Hereafter we refer to this choice of parameters $\wpara$ and $\wperp$ as $\WvecH$. \\
When measurement uncertainties are not known, a reasonable choice is to set 
\begin{eqnarray}
        w=
        \left\{
        \begin{array}{ll} 
        \wpara \\
        \wperp
        \end{array}
        \right\}
        = 
        \left\{
        \begin{array}{ll} 
        |\Bhor|/|\mathit{B_{\rm h,max}}|\\
        1.0
        \end{array}
        \right\}\label{eq:WvecE}
\end{eqnarray}
for each pixel separately. This choice was put forward by W12, based on a comparison of different definitions of $\Wvec$. With this particular choice  the vertical field is empirically measured at the  highest accuracy level, and that the accuracy of the measured horizontal field increases with its strength. Hereafter we refer to this choice of parameters $\wpara$ and $\wperp$  as $\WvecE$. To date, with the sole exception of M19, this setting has  often been applied when performing coronal NLFF modeling with the W12 method. This motivates us to test the performance of this type of error matrix, by comparison to the corresponding application of $\WvecH$.
\end{itemize}

Successful NLFF modeling involves  finding a combination of the free model parameters for the optimization function  $L$ and, if applied, for the preprocessing step ($\mu_3$, $\mu_4$) that delivers optimized results, in terms of force- and divergence-freeness. In order to quantify the force-free consistency of the obtained NLFF solutions for a certain choice of the free model parameters, a frequently used metric is the current-weighted angle between the modeled magnetic field and electric current density, $\thetaj$, \citep[][]{2006SoPh..235..161S}. Ideally, for an entirely force-free solution we would find $\thetaj=0\degree$.

As noted by \cite{2012SoPh..281...37W}, for the application to long-term HMI data series it is not practical to carry out NLFF modeling based on several different model parameter sets. For a short time span, as in our analysis, it is doable, and may be used to study the uncertainty of physical quantities based on the different model parameter choices.

\subsection{Helicity computation}\label{ss:fv_helicity}

We use the finite-volume (FV) method of \cite{2011SoPh..272..243T} to compute the relative helicity based on \href{eq:hv}{Eqs.~(\ref{eq:hv})}--\href{eq:hv}{(\ref{eq:hpj})}. It solves systems of partial differential equations to obtain the vector potentials $\Avec$ and $\Apvec$, using the Coulomb gauge, $\nabla\cdot\Avec=\nabla\cdot\Apvec=0$. The method defines the reference field as $\Bpvec=\nabla\phi$, with $\phi$ being the scalar potential, subject to the constraint $\nabla_n\phi=\Bvec_n$ on $\partial\mathcal{V}$, where $n$ denotes the normal component with respect to the boundaries of $\mathcal V$.

The method has been tested in the framework of an extended proof-of-concept study on FV helicity computation methods \citep[][]{2016SSRv..201..147V}, where it has been shown that for various test setups the methods deliver helicity values in line with each other, differing by only small percentage points. It has also been used in \cite{2019ApJ...880L...6T} to show that the computed helicity is highly dependent on the level to which the underlying NLFF magnetic field solution satisfies the divergence-free condition. A metric for quantifying the divergence-free consistency of an obtained NLFF solution, as introduced by \citep[][]{2000ApJ...540.1150W} and often used in literature is the fractional flux, $\fiavg$ \citep[for an recent in-depth analysis of this measure see][]{2020arXiv200808863G}. Though not shown explicitly in our work, we note for completeness that in all studied NLFF models, we find $\fiavg\lesssim4\times10^{-4}$.

Alternatively, in order to test the level of solenoidality of the magnetic field used as an input for helicity computation, the ratio $\Edivprime$ has been put forward by \citet{2013A&A...553A..38V} as a useful criterion. The value of $\Edivprime$ expresses the non-solenoidal fraction of the total (NLFF) energy. The quantity $\Ediv$ is derived from the solenoidal and non-solenoidal parts of the potential field ($\Bpvec=\Bps+\Bpns$) and current-carrying field ($\Bjvec=\Bjs+\Bjns$), which stem from the initial decomposition of the (NLFF) magnetic field into its potential ($\Bpvec$) and current-carrying ($\Bjvec$) component. Then the total energy of a given magnetic field may be written as the sum of the corresponding energy budgets in the form
\begin{eqnarray}
        E=\Eps+\Epns+\Ejs+\Ejns+\Emix, \label{eq:decomp_e}
\end{eqnarray}
with $\Emix$ being the energy corresponding to all cross terms \citep[see Eq.~(8) in][for details]{2013A&A...553A..38V}, and $\Ejs$ being a measure for the free energy. All contributions to $E$ in \href{eq:decomp_e}{Eq. (\ref{eq:decomp_e})}, except for $\Emix$, are positive definite. In the case that the input field is perfectly solenoidal,
we would find $\Epns=\Ejns=\Emix=0$, thus $\Ediv=0$. 
The energy associated with all non-solenoidal components can then be defined as 
\begin{eqnarray}
        \Ediv=\Epns+\Ejns+|\Emix|, \label{eq:ediv}
\end{eqnarray}
which represents an upper limit, as the absolute value of $\Emix$ is involved. Usually, $\Eps>\Ejs>\Emix>\Ejns>\Epns$ \citep[see, \eg,][]{2015ApJ...811..107D}. 

In the proof-of-concept study by \citet[][]{2016SSRv..201..147V}, based on solar-like numerical experiments, it was suggested that only $\Edivprime\lesssim0.08$ is sufficient for a reliable helicity computation. In a follow-up study, \cite{2019ApJ...880L...6T} suggested an even lower threshold ($\Edivprime\lesssim0.05$) for solar applications.

Based on the energy decomposition above, we can also use the non-solenoidal contribution to the free energy $\Emixprime$ to refine the quantification of the acceptable degree of non-solenoidality in an underlying NLFF model field. As shown in the comparative study of \cite{2015ApJ...811..107D}, the application of the W12 method, using standard choices for the (free) model parameters, may result in NLFF solutions with non-solenoidal errors on the order of the inherent free magnetic energy ($|\Emix|\simeq\Ejs$, see their Fig.~7b). It was also shown, however, that an alternative W12-based NLFF solution based on an adjusted set of model parameters (more precisely in setting $\wb>\wa$; see \href{sss:optim}{Sect.~\ref{sss:optim}}) resulted in a significant improvement of the solenoidal quality (see Appendix of their work). A refined quantification of the solenoidal quality of the NLFF magnetic fields in context with relative helicity computation, based on $\Emixprime$ has not been attempted so far.

\begin{table*}
 \tiny
\renewcommand{\arraystretch}{1.2}
\caption{Synoptic view model parameters for the employed NLFF models and their appearance in the manuscript (plot symbol and figure or appearance), if applicable.}             
\label{tab:cases}      
\centering                          
\begin{tabular}{l c c c c c l c}        
\hline\hline                 
Case & \multicolumn{4}{c}{Model parameters} & Appearance & Symbol &   Comment \\   
& \multicolumn{2}{c}{Preprocessing} & \multicolumn{2}{c}{Optimization} & (Figure) & &\\
& $\mu_3$ & $\mu_4$ & $\wb$ & $\Wvec$ & & &\\
\hline
\multicolumn{8}{c}{Special cases}\\
\hline
Sp1a & -- & --  & 1 & $\WvecH$ & 1 & \mycircle{SkyBlue} (light blue circle) & %
        No preprocessing applied. Input data is force-free inconsistent. \\
Sp1b &  -- `` -- &  -- `` --   & 1 &  $\WvecE$ & -- &  -- &  Not explicitly shown. Similar in behavior as Sp1a. \\
Sp1c &  -- `` -- &  -- `` --   &  2 &  $\WvecH$ & -- &  -- & -- `` --  \\
Sp1d &  -- `` -- &  -- `` --   &  2 &  $\WvecE$ & -- &  -- & -- `` --  \\
\multicolumn{8}{c}{\dotfill}\\
Sp2a & $0$ &  $0$  & 1 & $\WvecH$ & 1 & \mybullet{SkyBlue} (light blue bullet) & %
        Preprocessing applied.  Input data is force-free consistent.\\
Sp2b &  -- `` -- &  -- `` --   &  2 &  $\WvecH$ & -- &  -- & Not explicitly shown. Similar in behavior as Sp2a. \\
Sp2c &  -- `` -- &  -- `` --   & 1 &  $\WvecE$ & -- &  -- & -- `` --  \\
Sp2d &  -- `` -- &  -- `` --   &  2 &  $\WvecE$ & -- &  -- & -- `` --  \\
\multicolumn{8}{c}{\dotfill}\\
Sp3a & $0$ & $10^{-2}$  &  1 &  $\WvecH$  & -- & -- & 
        Preprocessing applied, including smoothing.\\
& & & & & & & Excluded from analysis due to insufficient solenoidal quality.\\
Sp3b &  -- `` -- &  -- `` --   &  2 &  $\WvecH$ & -- &  -- & -- `` --  \\
Sp3c &  -- `` -- &  -- `` --   & 1 &  $\WvecE$ & -- &  -- & -- `` --  \\
Sp3d &  -- `` -- &  -- `` --   &  2 &  $\WvecE$ & -- &  -- & -- `` --  \\
\multicolumn{8}{c}{\dotfill}\\
Sp4a & $10^{-3}$ & $0$  &  1 &  $\WvecH$   &  1, 3, 5a & \mybullet{NavyBlue} (dark blue bullet)& %
        Preprocessing applied, including nearness to observed data.\\
Sp4b &  -- `` -- & -- `` --  & 2 &  $\WvecH$ & 1, 3, 5a & \mysquare{NavyBlue} (dark blue square) &  %
        Same as Sp4a but $\wb=2$ used.\\
Sp4c & -- `` -- & -- `` --  & 1 & $\WvecE$ & 4, 5b & \mybullet{YellowOrange} (orange bullet) &  %
        Same as Sp4a but $\Wvec=\WvecE$ used.\\
Sp4d & -- `` -- & -- `` -- & 2 & $\WvecE$ & 4, 5b & \mysquare{YellowOrange} (orange square) &  %
        Same as Sp4c but $\wb=2$ used.\\
\hline
\multicolumn{8}{c}{Standard preprocessing cases} \\
\hline
St1a & $10^{-3}$ & $10^{-2}$ &  1 & $\WvecH$  & 1, 2, 3, 5a & \mybullet{Purple} (violet bullet) & %
        Preprocessing applied, including smoothing and nearness to observed data.\\
St1b & -- `` -- & -- `` --  &  2 & $\WvecH$ & 1, 2, 3, 5a & \mysquare{Purple} (violet square) &  %
        Same as St1a but $\wb=2$ used.\\
St1c & -- `` -- & -- `` --  & 1 & $\WvecE$ & 4, 5b & \mybullet{Red} (red bullet) &  %
        Same as St1a but $\Wvec=\WvecE$ used. (W12 default parameter set.)\\
St1d & -- `` -- & -- `` --  & 2 & $\WvecE$& 4, 5b & \mysquare{Red} (red square) & %
        Same as St1b but $\Wvec=\WvecE$ used.\\
\hline  
\end{tabular}
\end{table*}

\subsection{Choice of free model parameters}\label{ss:fparms}

The solenoidality of a NLFF solution obtained by minimizing $L$ in \href{eq:l_ff}{Eq.~(\ref{eq:l_ff})} naturally depends on $\Binput$ (thus, the free parameter choices of $\mu_3$ and $\mu_4$ in the preprocessing step, if applied), and on the choice of the diagonal elements of $\Wvec$ (either $\WvecH$ or $\WvecE$ in the present study). 

For any   combination of the above quantities, the divergence-freeness of the obtained NLFF solution can be enhanced by assigning a stronger relative importance of the divergence term, \ie, by choosing $\wb>\wa$ (see explanation of the free model parameters of $L$ in \href{sss:optim}{Sect.~\ref{sss:optim}} for details). As a consequence, a NLFF solution based on a certain choice of the other free parameters may not qualify to be used as an input to helicity computation when choosing $\wb=1$, but may do so when choosing an enhanced weight $\wb>\wa$. This has been found in \cite{2015ApJ...811..107D}, where the application of the standard setting $(\wa,\wb)=(1,1)$ was found to deliver an NLFF solution, but without  the solenoidal quality required for relative helicity computation, as the relative contribution of the mixed terms was comparable to that of the free energy ($|\Emix|\approx\Ejs$; see Fig.~7b in their work). The NLFF model solution based on the choice $(\wa,\wb)=(1,1.5)$, however, resulted in a significant decrease in the contribution $\Emix$ to the total energy, thus represented a valid input for relative helicity computation (with $|\Emix|<\Ejs$; see Fig.~11 in the Appendix of their work). In order to test the effect of an improved solenoidal quality of the NLFF model on the relative helicity computation, we therefore use the choices $(\wa,\wb)=(1,1)$ and $(\wa,\wb)=(1,2)$ in our work.

\href{tab:cases}{Table~\ref{tab:cases}} summarizes the tested NLFF time series regarding the specific parameter choices used to for their realization, their appearance throughout the manuscript, and the corresponding plotting symbols used. The influence of the distinct choice of preprocessing parameters has not been tested yet. Therefore, we use the specific combinations $(\mu_3,\mu_4)=(0,0)$, $(\mu_3,\mu_4)=(10^{-3},0)$, $(\mu_3,\mu_4)=(0,10^{-2})$, and $(\mu_3,\mu_4)=(10^{-3},10^{-2})$, the last representing the preferred standard choice suggested by W12 (``Standard preprocessing cases''). In combination with different combinations of optimization parameters $\wb=(1,2)$ and $\Wvec=(\WvecH,\WvecE)$, 16 NLFF model time series can be employed in total. Four additional  NLFF time series are employed based on the non-preprocessed (original) data. This is motivated by the fact that the minimization of the surface term in \href{eq:l_ff}{Eq.~(\ref{eq:l_ff})} might still yield a high-quality NLFF solution, despite the force-free incompatibility of the measured vector magnetic field. In this context, it should also be noted that the input magnetic field data (preprocessed or not) necessarily differs from the final NLFF lower boundary data due to the effect of the surface term. We note here that we do not explicitly show the results of all of the ``special  cases'' NLFF time series, as their relative performance is comparable to that of respective exemplary cases presented hereafter (see comments in the last column). Furthermore, we exclude all SP3 cases from the analysis, due to their insufficient solenoidal quality ($\Edivprime>0.1$).

As in M19, all NLFF computations were carried out for a computational volume of dimensions $320\times320\times320$ pixels, containing a buffer layer towards the lateral and top boundaries of $nd=32$~pixels, within which the NLFF solution drops in the form of a $\cos$-profile to the field prescribed on the boundaries (for details see W04). For relative helicity computation only the inner physical volume (excluding the buffer layer) was kept, and further cut in height at roughly two-thirds of the total height, yielding a size of the finally analyzed model coronal field of $256\times256\times203$ pixels.

\section{Results}

\subsection{Effect of preprocessing}\label{ss:prepro}

In order to test the effect of preprocessing, we minimize \href{eq:lppf}{Eq.~(\ref{eq:l_pp})} once using the standard relative weighting ($\mu_3=10^{-3}$, $\mu_4=10^{-2}$; hereafter ``standard preprocessing''), once omitting smoothing ($\mu_3=10^{-3}$, $\mu_4=0$), and once neglecting both ($\mu_3=\mu_4=0$). For the subsequent minimization of \href{eq:l_ff}{Eq.~(\ref{eq:l_ff})}, we apply the error matrix for optimization of the lower boundary as used in M19 ($\WvecH$) and use standard settings for the remaining model parameters, as suggested in W12 (see \href{sss:optim}{Sect.~\ref{sss:optim}}). As a kind of non-ideal reference we also run the optimization on the non-preprocessed data and compare the resulting NLFF time series in the following. 

\begin{figure}
        \resizebox{\hsize}{!}{\includegraphics{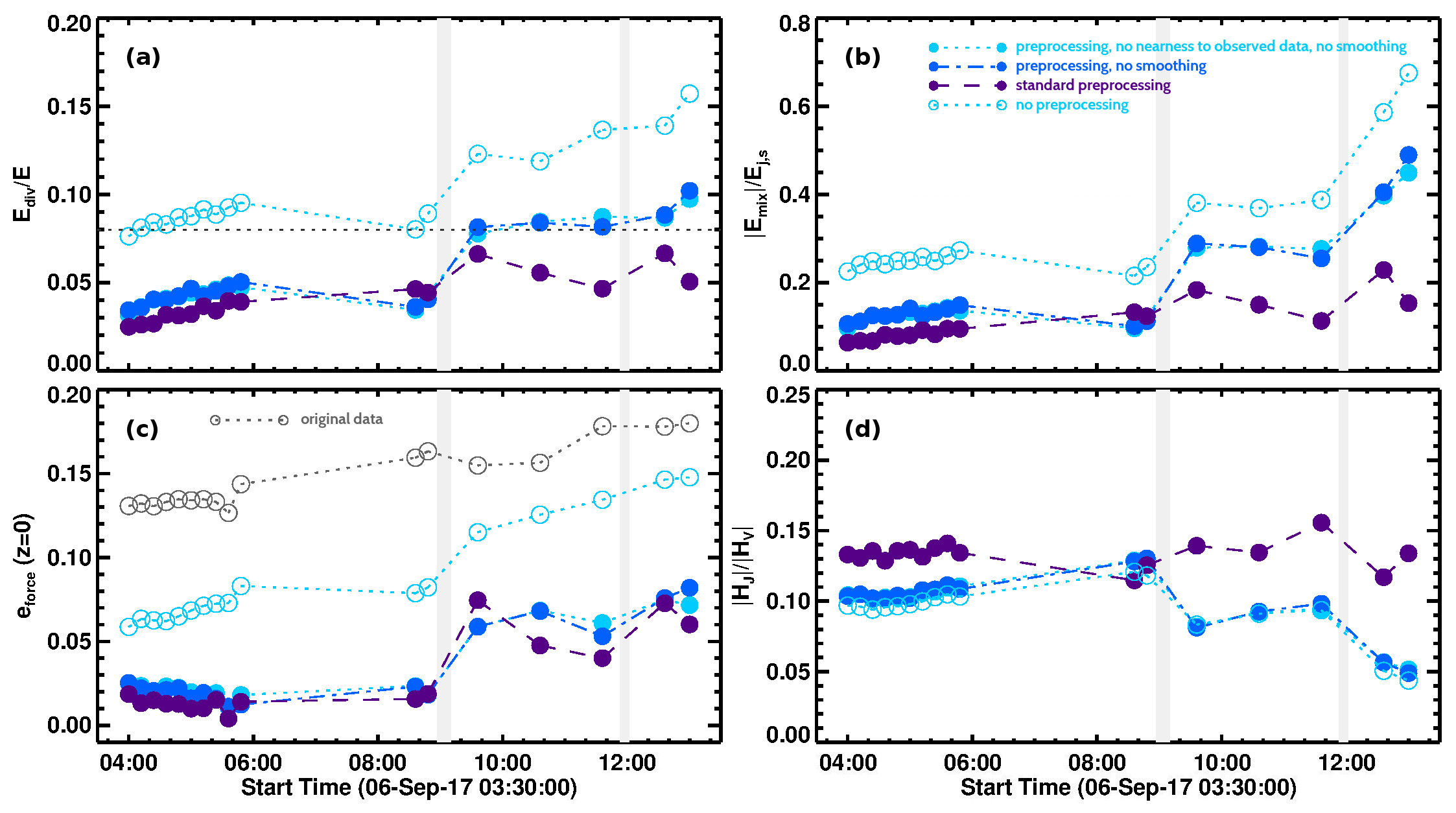}}
        \caption{Evolution of (a) $\Edivprime$ and (b) $\Emixprime$ and (c) $\efo$ for different NLFF models based on different input data: not preprocessed (light blue circles), standard preprocessed (violet bullets), preprocessed without smoothing (dark blue bullets), preprocessed without smoothing and no nearness to observed data enforced (light blue bullets). (d) Corresponding time evolution of $\hjprime$. The horizontal dashed line in (a) indicates the nominal threshold of $\Edivprime=0.08$, an upper limit for the accepted solenoidality of the input magnetic field.  The vertical bars show the time span between the nominal GOES start and peak time of the X2.2 (peak time 08:57~UT) and X9.3 flare (peak time 11:53~UT).}
        \label{fig:msk0_pp}
\end{figure}

From \href{fig:msk0_pp}{Fig.~\ref{fig:msk0_pp}a,b} it can be seen that the application of preprocessing clearly lowers the contribution of solenoidal errors. For the NLFF models based on non-preprocessed data (light blue circles) $\Edivprime$ and $\Emixprime$ are the highest at all considered times. Corresponding values are, on average, lowest for the NLFF solutions based on standard preprocessing (violet bullets). The effect of smoothing can be seen by comparison to the corresponding values of the ``no smoothing'' cases (dark and light blue bullets versus\ violet bullets). Overall, the application of smoothing causes a decrease in both $\Edivprime$ and $\Emixprime$, though more pronounced at later instances of the considered time period. It also appears that there is no distinct difference for the non-smoothed cases, whether or not enforcing a certain degree of nearness to the actually observed data (compare light and dark blue bullets). 

It is noteworthy that all NLFF time series show a deteriorating quality as a function of time, \ie, the values of $\Edivprime$ and $\Emixprime$ are increasing, supposedly due to the corresponding decrease in the inversion quality of the underlying vector magnetic field measurement (see \href{eq:WvecH}{Eq.~(\ref{eq:WvecH})} and corresponding notes), but is least pronounced for the NLFF solutions based on the standard preprocessed input. It is also noteworthy that all except those solutions exhibit values $\Edivprime\gtrsim0.08$ for the considered time instances after 08:48~UT.

In terms of $\thetaj$ the volumetric parameter usually used to quantify the force-free consistency of a NLFF model, there is no distinct difference between the cases when preprocessing is applied or not. For completeness we note that $\thetaj$ is below $\approx7\degree$ prior to the X2.2 flare and $7\degree\lesssim\thetaj\lesssim12\degree$ afterwards. In order to picture the effect of preprocessing more clearly, we therefore show the force-balance parameter, $\efo$, in \href{fig:msk0_pp}{Fig.~\ref{fig:msk0_pp}c}, which is normally used to quantify the force-free consistency of a given vector magnetogram prior to NLFF modeling \citep[see explanation in Sect.~2 of][]{2006SoPh..233..215W}. Here we use $\efo$ not only to quantify how force-free consistent the input data is, but also how force-free the final 2D NLFF lower boundary is. We know from previous studies that non-preprocessed vector magnetograph data is inconsistent with a force-free approach ($\efo\gtrsim0.1$; gray circles) and should not be used for force-free modeling. If used nevertheless, the optimization procedure will still deliver a NLFF solution, with its 2D lower boundary being more force-free consistent ($\efo\gtrsim0.08$; light blue circles). The application to preprocessed data clearly improves the NLFF model results ($\efo\lesssim0.08$; bullets) without any obvious dependencies on the particular parameter setting for preprocessing. 

The corresponding trends of $\hjprime$ (\href{fig:msk0_pp}{Figure~\ref{fig:msk0_pp}d}) suggest a clear segregation between NLFF time series based on smoothed (violet bullets) or non-smoothed (other symbols) input data. For completeness, we note that for all of the considered cases the relative helicities, $\hv$, based on smoothed data are systematically higher, as are their individual contributions (more pronounced in $\hj$ than in $\hpj$). We discuss the possible reasons in \href{s:summary}{Sect.~\ref{s:summary}}.

For completeness we note that the results presented above   also hold when using $\WvecE$ instead of $\WvecH$. For simplicity, and motivated by the the similarity of performance of the special cases ($\mu_3$,$\mu_4$)=(0,0) and  ($\mu_3$,$\mu_4$)=($10^{-3}$,0), respectively labeled  Sp2$i$ and Sp4$i$ in \href{tab:cases}{Table~\ref{tab:cases}}, we do not explicitly show the Sp2 cases in the remaining analysis.

\begin{figure}
        \resizebox{\hsize}{!}{\includegraphics{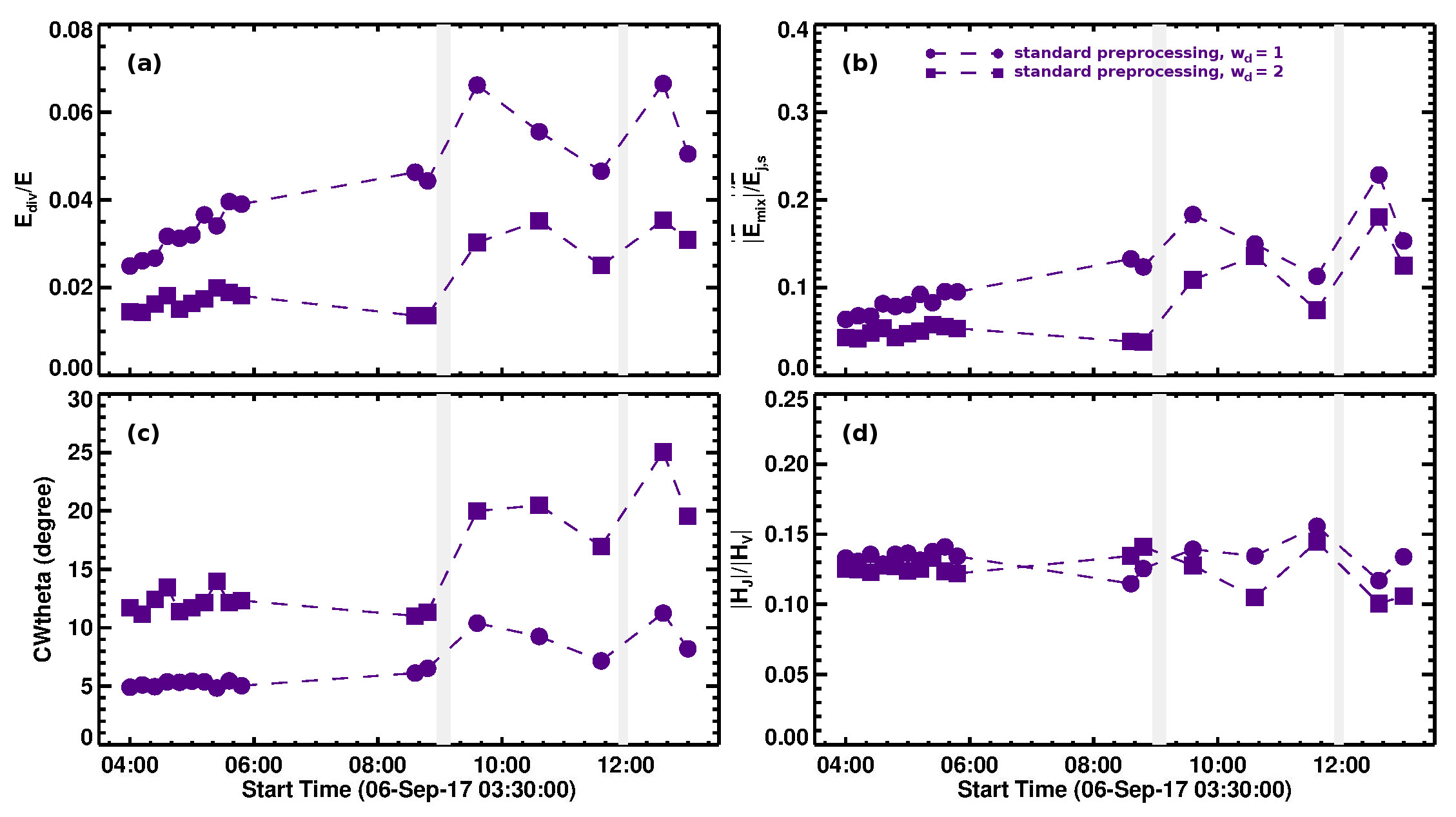}}
        \caption{Evolution of (a) $\Edivprime$ and (b) $\Emixprime$ and (c) $\thetaj$ for different NLFF models, based on standard preprocessed input data, with different weighting of the volume-integrated divergence: $\wb=1$ (bullets), and $\wb=2$ (squares); (d) Corresponding time evolution of $\hjprime$. Vertical bars as in \href{fig:msk0_pp}{Fig.~\ref{fig:msk0_pp}}.}
        \label{fig:msk0_wb}
\end{figure}

\subsection{Effect of preferring solenoidality over force-freeness}\label{ss:wb}

As explained in \href{ss:fparms}{Sect.~\ref{ss:fparms}}, for any choice of combination of other model parameters the divergence-freeness of the obtained NLFF solution may be enhanced by assigning a stronger relative importance of the divergence term, \ie, by choosing $\wb>\wa$. For simplicity, here we restrict ourselves to analyzing the corresponding effect based on standard preprocessed data (using $\mu_3=10^{-3}$ and $\mu_4=10^{-2}$ in \href{eq:l_pp}{Eq.~(\ref{eq:l_pp})}). For completeness, we note that the results presented in the following are similar for application to non-smoothed data ($\mu_4=0$; compare \href{ss:combined}{Sect.~\ref{ss:combined}} and \href{fig:msk0_combined}{Fig.~\ref{fig:msk0_combined}}), and also in the cases that the empirical error matrix, $\WvecE$, is used (see \href{ss:mask}{Sect.~\ref{ss:mask}} and \href{fig:msk2_combined}{Fig.~\ref{fig:msk2_combined}}).

In \href{fig:msk0_wb}{Fig.~\ref{fig:msk0_wb}} we compare the results based on the standard setting, where the Lorentz force and divergence term in \href{eq:l_ff}{Eq.~(\ref{eq:l_ff})} are weighted equally ($\wa=\wb=1$; bullets) to that with an enhanced enforcement of solenoidality ($\wb=2$; squares).  The stronger enforcement of solenoidality leads to lower values of $\Edivprime$ and $\Emixprime$ (\href{fig:msk0_wb}{Figs.~\ref{fig:msk0_wb}a} and \href{fig:msk0_wb}{\ref{fig:msk0_wb}b}, respectively), and is at the expense of the force-freeness of the obtained NLFF solutions (compare $\thetaj$ in \href{fig:msk0_wb}{Fig.~\ref{fig:msk0_wb}c}). Instead,  for the standard weighting, $\thetaj\lesssim10\degree$ for the entire time series, the values are about a factor of two higher if $\wb=2$ is applied. 

Both NLFF time series satisfy $\Edivprime<0.08$ (\href{fig:msk0_wb}{Fig.~\ref{fig:msk0_wb}a}), \ie, qualify for subsequent relative helicity computation. Though the obtained trend of $\hjprime$ in \href{fig:msk0_wb}{Fig.~\ref{fig:msk0_wb}d} is similar for most of the time instances, the NLFF series based on the standard setting ($\wb=1$; bullets) depicts a decrease  in $\hjprime$ prior to and an increase during the confined X2.2 flare, while the solutions based on $\wb=2$ (squares) indicate a pre-flare increase and subsequent helicity relaxation. Both time series agree on a helicity accumulation to values $\hjprime\gtrsim0.15$ prior to the eruptive X9.3 flare, and a pronounced helicity relaxation in correspondence to the eruptive flare.

\subsection{Combined effects}\label{ss:combined}

\begin{figure}
        \resizebox{\hsize}{!}{\includegraphics{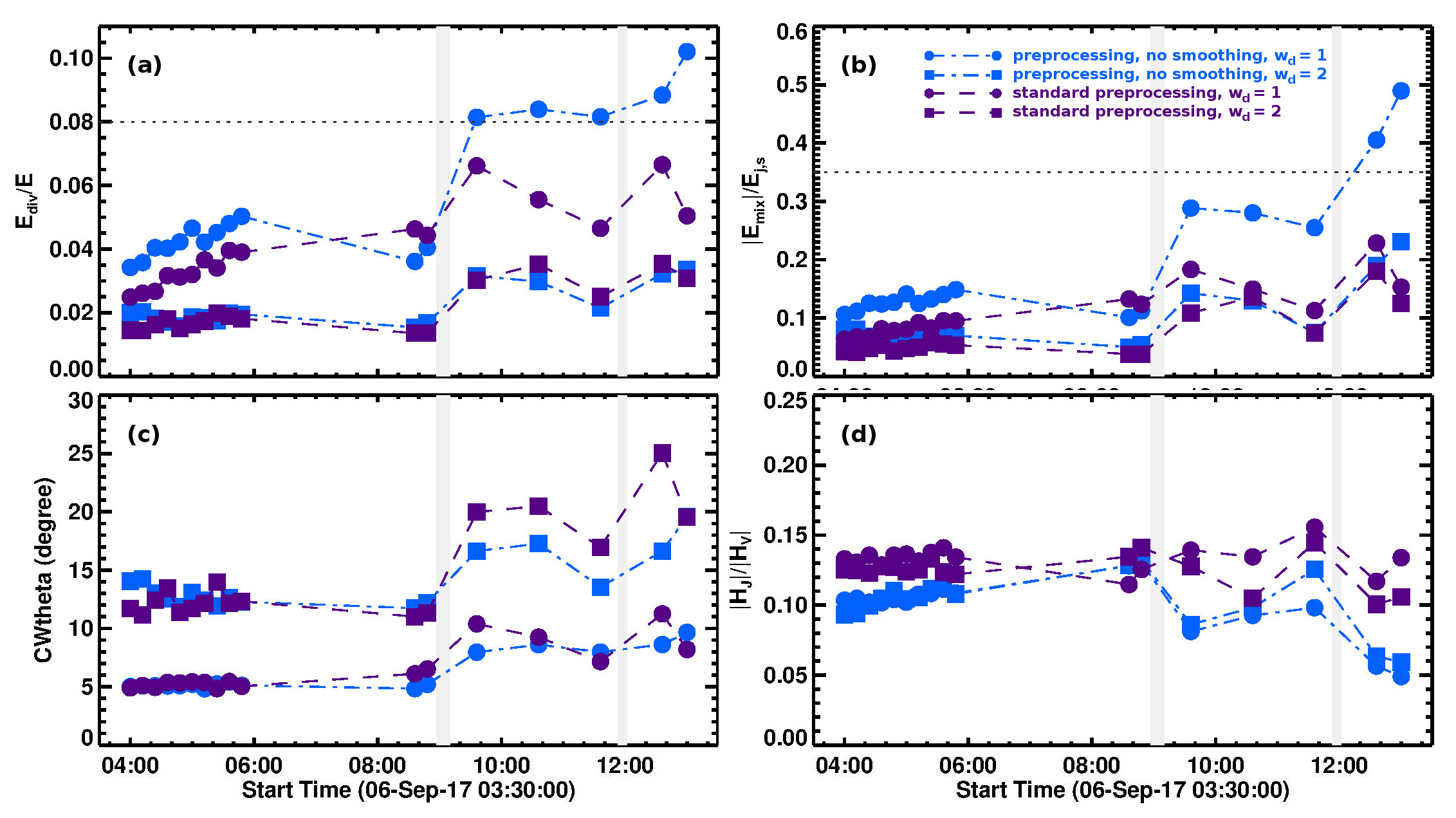}}
        \caption{Evolution of (a) $\Edivprime$ and (b) $\Emixprime$ and (c) $\thetaj$ for different NLFF models, 
        using the empirical error matrix $\WvecH$, and based on differently preprocessed input data (including smoothing: violet symbols; omitting smoothing: blue symbols), and with different weighting of the volume-integrated divergence ($\wb=1$: bullets; $\wb=2$: squares); (d) Corresponding time evolution of $\hjprime$. 
         The horizontal dashed line in (a) indicates the nominal threshold of $\Edivprime=0.08$, an upper limit for the accepted solenoidality of the input magnetic field. The horizontal dashed line in (b) indicates a suggested upper limit for the acceptable non-solenoidal error to the free magnetic energy ($\Emixprime=0.35$). Vertical bars as in \href{fig:msk0_pp}{Fig.~\ref{fig:msk0_pp}}.}
        \label{fig:msk0_combined}
\end{figure}
 
There is an interplay between particular choices of model parameters for the preprocessing and optimization, as individually discussed in \href{ss:prepro}{Sects.~\ref{ss:prepro}} and \href{ss:wb}{\ref{ss:wb}}, respectively. Therefore, we describe combined effects for the $\WvecH$-based models in the following, and those for the $\WvecE$-based models in \href{ss:mask}{Sect.~\ref{ss:mask}}. 

The choice $\wb=2$ (squares) during optimization has a similar effect on the final NLFF solution, regardless of whether smoothed (violet symbols) or non-smoothed (blue symbols) input data are used. On average, this value lowers the non-solenoidal energy contributions (\href{fig:msk0_combined}{Figs.~\ref{fig:msk0_combined}a} and \href{fig:msk0_combined}{\ref{fig:msk0_combined}b}) and simultaneously increases $\thetaj$ to a comparable level (compare \href{fig:msk0_combined}{Fig.~\ref{fig:msk0_combined}c}). The effective increase in solenoidality, however, is stronger for the NLFF models based on non-smoothed data (blue symbols), yielding the improved NLFF series to fulfill $\Edivprime<0.08$ at all times (blue squares). It also appears that NLFF solutions with a smaller value of $\Edivprime$ also exhibit on overall smaller values of $\Emixprime$.

It is also evident that NLFF series of comparable solenoidal quality do not necessarily deliver similar helicity ratios (compare  blue and violet squares in \href{fig:msk0_combined}{Figs.~\ref{fig:msk0_combined}b} and \href{fig:msk0_combined}{\ref{fig:msk0_combined}d}). Instead, the values of $\hjprime$ retrieved from non-smoothed boundaries (blue symbols) are found to be  systematically lower for most time instances. Even so,  all of the tested solutions depict a decrease in $\hjprime$ during both flares with the sole exception of the NLFF solutions based on the standard preprocessed data (violet bullets), which suggest a decrease in $\hjprime$ during the preceding confined X2.2 flare. All of the tested solutions show a rise in $\hjprime$ prior to the X9.3 flare to values close to those prior to the X2.2 flare.

\begin{figure}
        \resizebox{\hsize}{!}{\includegraphics{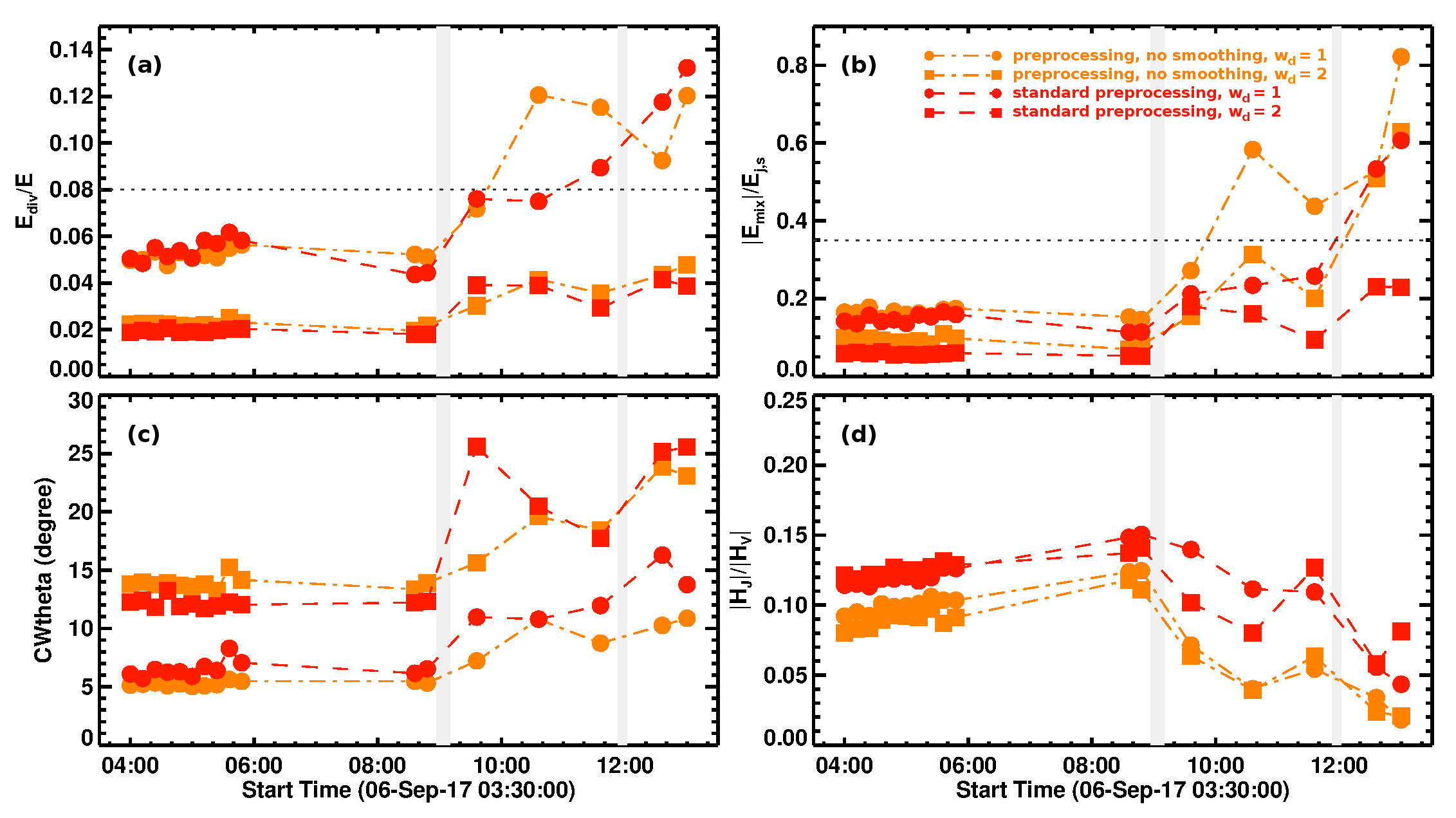}}
        \caption{As in \href{fig:msk0_combined}{Fig.~\ref{fig:msk0_combined}}, but using the empirical error matrix $\WvecE$.}
        \label{fig:msk2_combined}
\end{figure}

\subsection{Effect of the particular choice of error matrix $\Wvec$}\label{ss:mask}

As noted by W12, a reasonable approximation of the accuracy of the measured vector magnetogram data may be such that it weights the vertical magnetic field measurement strongest (based on its empirically known highest measurement accuracy), followed by the strong horizontal field, and with weak horizontal fields being weighted least strong (see $\WvecE$ as defined in \href{eq:WvecE}{Eq.~(\ref{eq:WvecE})} in \href{sss:optim}{Sect.~\ref{sss:optim}}). In the following we test the performance of this empirical weighting, and repeat the model experiments applied to the measurement-based error matrix $\WvecH$ as presented in \href{ss:combined}{Sect.~\ref{ss:combined}}.

Trends common to those presented in \href{ss:combined}{Sect.~\ref{ss:combined}} for the $\WvecH$-based models include that the choice $\wb=2$ (squares) during optimization lowers the non-solenoidal energy contributions (\href{fig:msk2_combined}{Figs.~\ref{fig:msk2_combined}a} and \href{fig:msk2_combined}{\ref{fig:msk2_combined}b}), while $\thetaj$ is systematically higher (\href{fig:msk2_combined}{Fig.~\ref{fig:msk2_combined}c}). In addition, systematically lower values of $\hjprime$ are found for all time instances from the NLFF models that are based on non-smoothed input data (orange symbols). For the $\WvecE$-based models the individual relative helicity contributions, $\hpj$ and especially $\hj$, are also systematically higher based on non-smoothed input data, and more pronounced than for the $\WvecH$-based models. The increase in $\hjprime$ between the two consecutive flares is less pronounced than for the $\WvecH$-based models. Finally, all of the tested $\WvecE$-based models consistently picture a decrease in $\hjprime$ during both X-class flares, as well as periods of helicity accumulation prior to both flares (though rather weak for most of the NLFF series).

Other findings are different from those of the $\WvecH$-based models. For instance, the $\WvecE$-based solutions show a deteriorating quality as a function of time, more pronounced than the $\WvecH$-based solutions (compare \href{fig:msk2_combined}{Figs.~\ref{fig:msk2_combined}a,b} and \href{fig:msk0_combined}{\ref{fig:msk0_combined}a,b}). Furthermore, a lower value of $\Edivprime$ does not necessarily imply a lower value of $\Emixprime$ for the $\WvecE$-based models, and thus NLFF solutions that satisfy $\Edivprime<0.08$ may exhibit large non-solenoidal contributions to the free magnetic energy.

\subsection{Putting everything together: a recipe}

Not all of the analyzed NLFF models presented in \href{ss:combined}{Sects.~\ref{ss:combined}} and \href{ss:mask}{\ref{ss:mask}} satisfy the nominal threshold of $\Edivprime<0.08$ \citep[as suggested by ][]{2016SSRv..201..147V} at all considered times, \ie, not all do qualify for relative helicity computation. Besides, based on the time evolution of the relative helicity in NOAA~11158, computed with different FV helicity computation methods, \cite{2019ApJ...880L...6T} argued for {an even more restrictive threshold in solar applications ($\Edivprime\lesssim0.05$). This is approximately fulfilled for the $\WvecH$-based models that satisfy the nominal threshold $\Edivprime<0.08$, inherently including a small non-solenoidal contribution to the free magnetic energy ($\Emixprime\lesssim0.25$).

For the $\WvecE$-based models, however, values of $\Edivprime<0.08$ do not necessarily imply small values of $\Emixprime$ (see orange and red squares at the last two time instances in \href{fig:msk2_combined}{Fig.~\ref{fig:msk2_combined}b}, for which $\Emixprime\gtrsim0.5$). Thus, NLFF solutions with high levels of non-solenoidal energy compared to their free energy would enter the relative helicity computation. As mentioned before, it appears crucial for applications of the W12 method to minimize the non-solenoidal contribution to the free magnetic energy (see corresponding remarks in \href{ss:fv_helicity}{Sect.~\ref{ss:fv_helicity}}). We thus suggest, in addition to respecting the nominal threshold of $\Edivprime=0.08$, to keep only the best-performing snapshots in terms of smallest non-solenoidal contribution to the free magnetic energy from each NLFF time series, namely those for which $\Emixprime<0.35$.

In order to place the contribution of $\Emix$ into context, we note here that the free magnetic energy during the analyzed time interval is in the range $\gtrsim20$\% of the total magnetic energy (\ie, $\Ejsprime\gtrsim0.2$). Since $\Emix$ comprises a small percentage of $\Etot$ only ($\lesssim10$\%), we may safely assume that it is rooted in numerical reasons, and that a corresponding thresholding has the desired effect of  sorting out NLFF solutions with a related undesirably high contribution.

Based on the above reasoning, we can then compute a mean value, $\langle\hjprime\rangle$, from all of the accepted NLFF solutions at each time instant, and also deduce an uncertainty estimate based on the spread of the contributing solutions. For time instances when only one contributing NLFF solution remains based on the above selection criteria, no mean value can be retrieved and the respective value of $\hjprime$ has to be assumed as indicative of the true coronal relative helicity.

\href{fig:msk0_msk2_ejmix}{Figure~\ref{fig:msk0_msk2_ejmix}a} shows the time evolutions of $\hjprime$, computed from all qualifying $\WvecH$-based NLFF models (colored symbols), together with the mean value $\mhjprime$ (black solid line) and the standard deviation as an indication of the corresponding uncertainty (gray shaded area). The corresponding time evolutions for the $\WvecE$-based NLFF models are shown in \href{fig:msk0_msk2_ejmix}{Figure~\ref{fig:msk0_msk2_ejmix}b}. Overall, the $\WvecH$-based estimates show less variation of $\mhjprime$ as a function of time, though the trend is very similar to that of the $\WvecE$-based estimates. A period of rather monotonic helicity accumulation prior to the confined X2.2 flare terminates in values of $\mhjprime\simeq0.13$. The subsequent flare-related helicity relaxation (apparently more pronounced in the $\WvecE$-based estimates) is followed by a period of relative helicity replenishment between $\sim$10:00~UT and 11:30~UT, peaking shortly before the eruptive X9.3 flare ($\mhjprime\gtrsim0.12$). Finally, the X9.3 flare-related helicity relaxation shows a decrease to values of $\mhjprime\lesssim0.1$.

\begin{figure}
        \resizebox{\hsize}{!}{\includegraphics{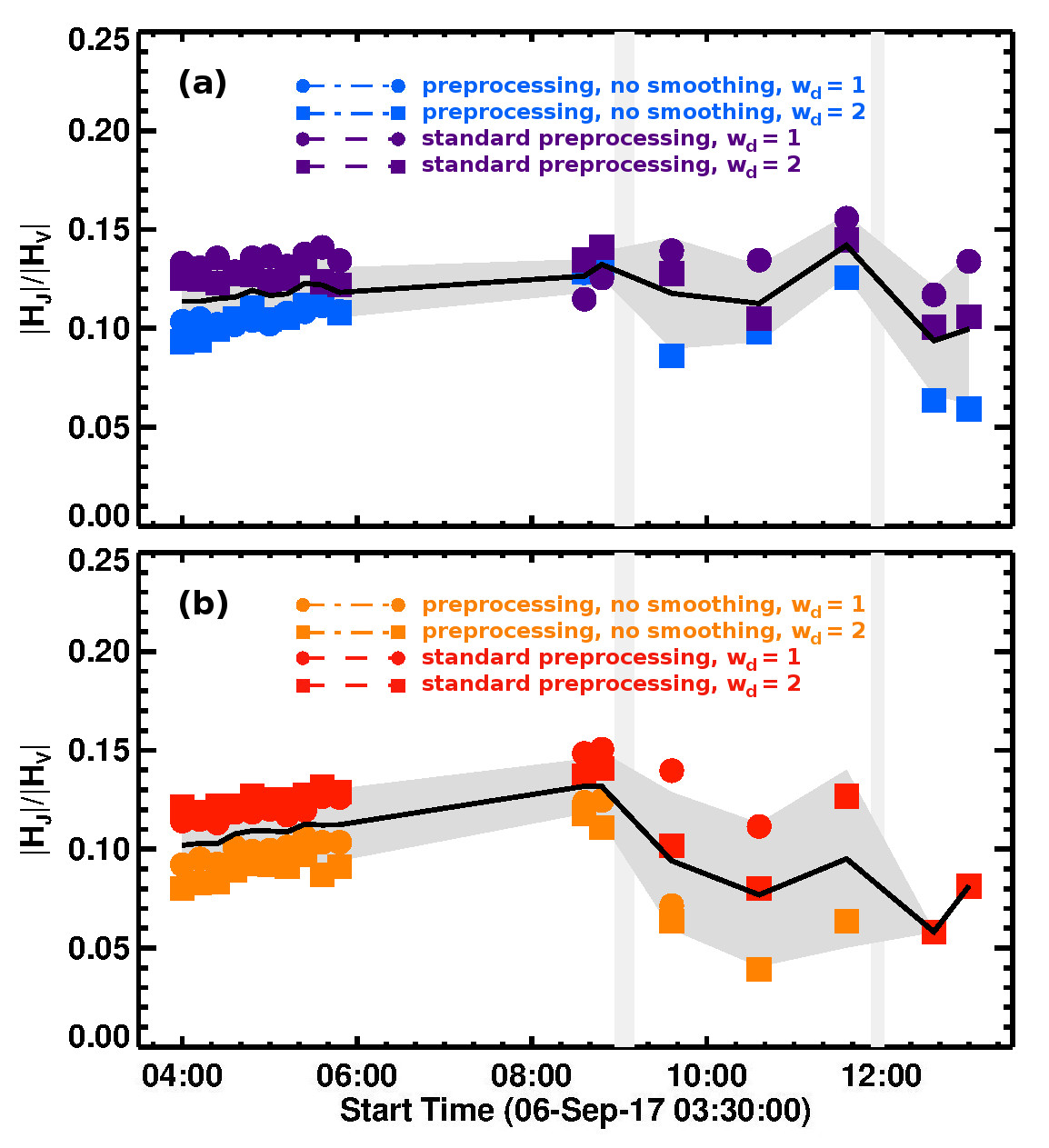}}
        \caption{Time evolution of $\hjprime$, computed from the best-performing NLFF models using the error matrix (a) $\WvecH$ or (b) $\WvecE$. At each time instance, only contributions from NLFF solutions are shown that satisfy $\Emixprime\leq0.4$. The black solid line represents the mean value, computed from all qualifying solutions at each time instant. The gray shaded area represents the respective standard deviation. Vertical bars as in \href{fig:msk0_pp}{Fig.~\ref{fig:msk0_pp}}.}
        \label{fig:msk0_msk2_ejmix}
\end{figure}

\section{Summary}\label{s:summary}

We studied the coronal magnetic field and helicity of AR~12673, a ten-hour time interval centered around a preceding X2.2 flare (SOL2017-09-06T08:57) that also includes an eruptive X9.3 flare that occurred three hours later (SOL2017-09-06T11:53). Our aim was to assess the spread of the relative helicities computed from \href{eq:hv}{Eqs.~(\ref{eq:hv})} and \href{eq:hpj}{(\ref{eq:hpj})}, using the finite-volume (FV) method of \cite{2011SoPh..272..243T} when based on different time series of NLFF coronal magnetic fields. The corresponding NLFF coronal magnetic fields were modeled using an optimization approach \citep[][hereafter W12]{2012SoPh..281...37W} based on different choices of (free) model parameters, consequently differing in their solenoidal quality. The solenoidal quality of an underlying NLFF model, however, is a highly important factor if one attempts a reliable relative helicity computation, and different thresholds have been suggested in the past \citep{2016SSRv..201..147V,2019ApJ...880L...6T}. 

The aim of our study was to  gain insight into the effects of particular choices of (free) model parameters on the final W12 NLFF solutions and subsequent relative helicity computation. Based on the in-depth analysis of the solenoidal quality of the underlying NLFF solutions, in context with the subsequently computed relative helicity ratio, $\hjprime$ \citep[a promising indicator for the eruptivity of solar ARs; see][for a pioneering work]{2017A&A...601A.125P}, our goal was to provide a recipe for successful and reliable relative helicity computation (including uncertainties).

The W12 method involves two computational tasks. In the first preprocessing step, individual weights can be assigned to the nearness to the actually observed data and the degree of smoothing applied to the 2D vector magnetic field data (controlled by $\mu_3$ and $\mu_4$, respectively, see \href{sss:prepro}{Sect.~\ref{sss:prepro}} for details). In our work we tested  the standard choices $\mu_3=10^{-3}$ and $\mu_4=10^{-2}$, and also the limiting values $\mu_3=0$ and $\mu_4=0$. In the subsequent optimization step (see \href{sss:optim}{Sect.~\ref{sss:optim}} for details), the volume-integrated Lorentz force and divergence can be weighted individually (via $\wa$ and $\wb$, respectively), and the handling of the (preprocessed) vector magnetic field data can be defined. The latter is realized by a diagonal error matrix where we tested two options. First, an error matrix originally defined in the study of \cite{2019A&A...628A..50M} (hereafter M19), based on the actual measurement uncertainties of HMI ($\WvecH$ as defined in \href{eq:WvecH}{Eq.~(\ref{eq:WvecH})}), and then a commonly used empirical one ($\WvecE$ as defined in \href{eq:WvecE}{Eq.~(\ref{eq:WvecE})}; see also W12) which assumes that vertical fields are measured with highest accuracy, and that the reliability of the measured horizontal field decreases with decreasing strength.

The volumetric force-freeness of the realized NLFF models was estimated using the current-weighted angle between the modeled magnetic field and electric current density, $\thetaj$, \citep[][]{2006SoPh..235..161S}. For the quantification of their solenoidal quality, we used the normalized non-solenoidal energy ratio $\Edivprime$, as suggested by \citep{2013A&A...553A..38V}. Moreover, since it appears crucial to minimize the non-solenoidal contribution to the free magnetic energy (see corresponding remarks in \href{ss:fv_helicity}{Sect.~\ref{ss:fv_helicity}}), we analyzed the energy ratio $\Emixprime$ in detail in our work, which has not been studied before.

Regarding the impact of particular choices of (free) model parameters, independent of the particular error matrix used ($\WvecH$ or $\WvecE$), on the solenoidal quality of the final NLFF solutions we found the following:

\begin{itemize}
        \item[--] The application of preprocessing prior to optimization considerably lowers the non-solenoidal contributions in the final NLFF solution, with the NLFF solution being also more force-free (\href{fig:msk0_pp}{Fig.~\ref{fig:msk0_pp}a--c}). \\
        A crucial ingredient in lowering the solenoidal errors appears to be the application of smoothing ($\mu_4\neq0$). Thus, the solenoidal quality of NLFF solutions based on the preprocessing as suggested in W12 ($\mu_1=\mu_2=1$, $\mu_3=10^{-3}$, $\mu_4=10^{-2}$) is   highest overall  and can be safely recommended as a standard setting.
        \item[--] The enhanced weighting of the volume-integrated divergence over the force-freeness ($\wb>\wa$) also lowers the non-solenoidal contributions in the final NLFF solution (\href{fig:msk0_wb}{Figs.~\ref{fig:msk0_wb}a} and \href{fig:msk0_wb}{\ref{fig:msk0_wb}b}), though at the expense of force-freeness (compare \href{fig:msk0_wb}{Fig.~\ref{fig:msk0_wb}c}).\\ 
        The effective increase in solenoidal quality is more drastic for the NLFF models based on non-smoothed data (blue and orange symbols in \href{fig:msk0_combined}{Figs.~\ref{fig:msk0_combined}a,b} and \href{fig:msk2_combined}{\ref{fig:msk2_combined}a,b}, respectively), underlining the corresponding desired effect of using smoothed data as input to NLFF modeling.
\end{itemize}

Different choices of the (free) model parameters during preprocessing and optimization allow the computation of multiple values for the relative helicities ($\hv$, $\hpj$, $\hj$), and consequently for the helicity ratio, $\hjprime$, at a certain time instant for a particular error matrix (for $\WvecH$- and $\WvecE$-based models see \href{fig:msk0_combined}{Fig.~\ref{fig:msk0_combined}d} and \href{fig:msk2_combined}{Fig.~\ref{fig:msk2_combined}d}, respectively). In this context, we found the following causal impacts:
\begin{itemize}
        \item[--] 
        Using smoothed data as input to NLFF modeling yields systematically higher values of $\hjprime$ (blue and orange symbols in \href{fig:msk0_combined}{Figs.~\ref{fig:msk0_combined}d} and \href{fig:msk2_combined}{\ref{fig:msk2_combined}d}, respectively), and similarly for the individual contributions, $\hpj$ and $\hj$, (independent of the error matrix used). \\
        Although $\hj$ has a clear physical meaning, namely the linking of the current-carrying field with itself, an enhanced level of $\hj$ does not necessarily imply the presence of systematically stronger electric currents  \citep{2009A&A...497L..17R}. And indeed, we do not find a systematically higher total unsigned current in the NLFF models based on smoothed data. Instead, we find higher total magnetic energies, $\Etot$, and lower potential field energies, $\Epot$, in those models. Thus, we suspect that the origin of the  overall higher helicities for the NLFF models is based on smoothed data in the systematically enhanced current-carrying magnetic field.
        \item[--] For those NLFF models that satisfy the originally suggested threshold of $\Edivprime<0.08$, the non-solenoidal contributions to the free energy, $\Emixprime$, are distinctly different. While the $\WvecH$-based NLFF solutions satisfying $\Edivprime\lesssim0.05$ \citep[a refined threshold for solar applications suggested by][]{2019ApJ...880L...6T} also satisfy $\Emixprime\lesssim0.25$, this is not true for the $\WvecE$-based models.\\
        Therefore, and motivated by minimizing non-solenoidal errors in the free magnetic energy, an additional threshold based on $\Emixprime$ appears useful in order to (dis-)qualify NLFF solutions for subsequent relative helicity computation.
        \item[--] Using an upper limit of $\Emixprime=0.35$, we obtain similar trends for the mean time evolution, $\langle\hjprime\rangle$, from the two types of NLFF series (based on either $\WvecH$ or $\WvecE$, see \href{fig:msk0_msk2_ejmix}{Fig.~\ref{fig:msk0_msk2_ejmix}a} and \href{fig:msk0_msk2_ejmix}{\ref{fig:msk0_msk2_ejmix}b}, respectively). Then, the empirical error matrix $\WvecE$ may be validly used as an alternative to a measurement-based definition (such as $\WvecH$).
\end{itemize}

Based on the above findings, we are able to provide a recipe to obtain a reliable estimate of the coronal relative helicity together with a corresponding uncertainty estimate. In particular, we recommend  employing a mean estimate of the relative helicity (and of any related quantity such as $\langle\hjprime\rangle$) at any particular time instant, computed from a number of NLFF models based on different (free) model parameter choices that individually satisfy $\Edivprime<0.08$ and $\Emixprime<0.35$. Using this approach we found a consistent estimate of $\mhjprime$ from the two types of NLFF model series ($\WvecE$- and $\WvecH$-based). This includes an increase in $\mhjprime$ prior to the confined X2.2 flare and between the preceding X2.2 and following X9.3 flare, together with helicity (ratio) relaxation in correspondence to the occurrences of the  flares. 

However, the spread of the contributing values of $\hjprime$ is quite variable over the time series, about $\lesssim0.04$   before the occurrence of the X2.2 flare and $\gtrsim0.06$ prior to the eruptive flare. Overall it appears that the spread of $\hjprime$ scales with the quality of the underlying NLFF time series. We recall  that all of the employed NLFF time series show a deteriorating quality, \ie, the values of $\Edivprime$ and $\Emixprime$  increase with time, supposedly due to the corresponding decrease in the inversion quality of the underlying vector magnetic field measurement (see corresponding notes in \href{sss:optim}{Sect.~\ref{sss:optim}}).

\section{Discussion}
 
Multiple attempts were made  to model and interpret the coronal magnetic field configuration of AR~12673,  focused on the approximation of the self-helicity of a coronal model flux rope recovered from NLFF modeling. We note here for completeness that in all of the finally qualifying NLFF time series, a magnetic flux rope is present prior to the confined X2.2 flare, of differing morphology but in overall agreement with earlier model attempts. Therefore, we assume that our NLFF model fields   realistically represent the active-region corona of AR~12673. An in-depth comparison of the distinct model magnetic field configurations, including the extent of recovering a possibly existing ``double-decker'' system (see discussion below), is left for a future work.

Based on an magneto-hydrodynamic relaxation method, \cite{2020ApJ...890...10Z} pictured the formation and gradual growing of a magnetic flux rope prior to the confined X2.2 flare, covering the time span 00:00~UT to 11:48~UT on 2017~September~6. The existing magnetic flux rope was found to grow in an accelerated manner after the occurrence of the confined flare, along with a (mild) increase in the flux rope twist (an approximation for its self-helicity). In agreement, though not explicitly shown, we note that all our tested NLFF model series depict rather monotonically increasing relative helicity $\hv$ (and its individual contributions $\hpj$ and $\hj$) before the X2.2 flare, and show another increase  between the preceding X2.2 and following X9.3 flare. Based on a series of optimization-based NLFF models, \cite{2018ApJ...867L...5L} pictured the pre-X2.2 flare coronal magnetic field configuration in the form of a system of multiple flux ropes, overlying each other and composed of field of opposite handedness \citep[double-decker; see also][]{2018A&A...619A.100H}. In particular, using the twist number method, they pictured a considerable increase in the flux rope  twist during the confined flare. Based on the same method, \cite{2020ApJ...890...10Z} pictured a rather weakly increasing self-helicity during the X2.2 flare (see their Fig.~4c). In this context we note that only one of our tested NLFF model series depicts a weak increase in $\hjprime$ during the X2.2 flare (violet bullets in \href{fig:msk0_wb}{Fig.~\ref{fig:msk0_wb}d}). All the other NLFF series, and consequently $\mhjprime$, depict a corresponding relative helicity relaxation. This does not necessarily conflict with a flare's confined nature, since a corresponding variation in $\hj$ may be simply due to the exchange with $\hpj$ \citep{2018ApJ...865...52L}. Finally, all of our tested NLFF series suggest a relative helicity relaxation (and also of $\mhjprime$) during the eruptive X9.3 flare. In agreement with  \cite{2018ApJ...867L...5L}, among others,  this is expected since the current-carrying magnetic structure (\ie, the coronal flux rope) was physically ejected from the corona.

In M19 the relative helicity of NOAA~12673 was studied in detail, based on a mix of NLFF models computed using either the W04 or W12 method (using standard model parameter choices), depending on which of the NLFF fields had a lower value $\Edivprime$, and necessarily $\Edivprime<0.08$ (see their Fig.~4). In particular, the W12 models at 08:36~UT and 08:48~UT were of lower solenoidal quality than the corresponding W04 solutions and were  thus dropped from the analysis. For the remaining time instances the W12 solutions were retained due to their relatively lower solenoidal errors. The resulting time evolution of $\hjprime$ depicted an increase in $\hjprime$ to values $\gtrsim0.15$ prior to the X-class flares, corresponding decreases in the course of the flares, and the replenishment of relative the helicity ratio before the X9.3 flare (see their Fig.~7). In comparison, we find a similar time evolution of $\hjprime$, though indicating lower characteristic pre-flare values of $\simeq0.13$ (violet bullets in \href{fig:msk0_wb}{Fig.~\ref{fig:msk0_wb}d}). Only the pronounced pre-X2.2 flare peak of $\hjprime>0.15$ found in M19 is not recovered in our NLFF solutions. It should be noted that  their estimates at 08:36~UT and 08:48~UT were based on two W04-based solutions with values of $\Edivprime$ marginally below the nominal threshold of $\Ediv=0.08$.   For NLFF models with $\Edivprime\gtrsim0.05$, however, estimates of $\hjprime$ may vary considerably among different helicity computation methods, even when based on the same sequence of NLFF models \citep[compare Figs.~2c and 4c in][]{2019ApJ...880L...6T}. We therefore may explain our lower pre-X2.2 flare values as the inherent uncertainty of relative helicity estimates for a solenoidal quality of the underlying NLFF models in the regime $0.05\lesssim\Edivprime\lesssim0.08$.

\section{Conclusion}

In conclusion, reliable estimations of the relative helicity budget (and that of related quantities) based on NLFF coronal magnetic field models remain  a challenging task. The extended analysis of the various NLFF model parameters in this work and the comparison with the analysis presented in M19 showed that finite-volume relative helicity computation is highly sensitive to the details of the underlying magnetic field modeling.

A way to compensate for related issues is to employ multiple NLFF time series based on different (free) model parameter choices and to employ mean estimates based on the subset of NLFF models that satisfy $\Edivprime<0.08$ and $\Emixprime<0.35$ at a particular time instant. In this way, it is possible to obtain reliable estimates of the relative helicity (and related quantities) along with corresponding uncertainty estimates. This  involves a large computational effort and time, but it substantially increases  the understanding and the reliability of the obtained results. As noted by W12, this might not be doable for long time series, but might be a favorable approach around the times of the occurring flares.

\begin{acknowledgements} 
We thank the anonymous referee for careful consideration of the manuscript and insightful comments.  J.\,K.\,T.\ and M.\, G.\ were supported by the Austrian Science Fund (FWF): P31413-N27. X.\,S.\ is partially supported by NSF awards \#1848250, \#1854760, and NASA award \#80NSSC190263. K.\,M.\ acknowledges support of the French Agence Nationale pour la Recherche through the HELISOL project ANR-15-CE31-0001. {\it SDO} data are courtesy of the NASA/{\it SDO} AIA and HMI science teams. This article profited from discussions during the meetings of the ISSI International Team {\it Magnetic Helicity in Astrophysical Plasmas}.
\end{acknowledgements}

\bibliographystyle{aa} 

\begin{thebibliography}{41}
\expandafter\ifx\csname natexlab\endcsname\relax\def\natexlab#1{#1}\fi

\bibitem[{{Aly}(1989)}]{1989SoPh..120...19A}
{Aly}, J.~J. 1989, \solphys, 120, 19

\bibitem[{{Berger}(1999)}]{1999PPCF...41B.167B}
{Berger}, M.~A. 1999, Plasma Physics and Controlled Fusion, 41, B167

\bibitem[{{Berger} \& {Field}(1984)}]{1984JFM...147..133B}
{Berger}, M.~A. \& {Field}, G.~B. 1984, Journal of Fluid Mechanics, 147, 133

\bibitem[{{DeRosa} {et~al.}(2015){DeRosa}, {Wheatland}, {Leka}, {Barnes},
  {Amari}, {Canou}, {Gilchrist}, {Thalmann}, {Valori}, {Wiegelmann},
  {Schrijver}, {Malanushenko}, {Sun}, \& {R{\'e}gnier}}]{2015ApJ...811..107D}
{DeRosa}, M.~L., {Wheatland}, M.~S., {Leka}, K.~D., {et~al.} 2015, \apj, 811,
  107

\bibitem[{{Finn} \& {Antonsen}(1984)}]{1984CPPCF...9..111F}
{Finn}, J. \& {Antonsen}, T.~J. 1984, Comments Plasma Phys. Controlled Fusion,
  9, 111

\bibitem[{{Gilchrist} {et~al.}(2020){Gilchrist}, {Leka}, {Barnes}, {Wheatland
  }, \& {DeRosa}}]{2020arXiv200808863G}
{Gilchrist}, S.~A., {Leka}, K.~D., {Barnes}, G., {Wheatland }, M.~S., \&
  {DeRosa}, M.~L. 2020, arXiv e-prints, arXiv:2008.08863

\bibitem[{{Hou} {et~al.}(2018){Hou}, {Zhang}, {Li}, {Yang}, \&
  {Li}}]{2018A&A...619A.100H}
{Hou}, Y.~J., {Zhang}, J., {Li}, T., {Yang}, S.~H., \& {Li}, X.~H. 2018, \aap,
  619, A100

\bibitem[{{James} {et~al.}(2018){James}, {Valori}, {Green}, {Liu}, {Cheung},
  {Guo}, \& {van Driel-Gesztelyi}}]{2018ApJ...855L..16J}
{James}, A.~W., {Valori}, G., {Green}, L.~M., {et~al.} 2018, \apj, 855, L16

\bibitem[{{Linan} {et~al.}(2018){Linan}, {Pariat}, {Moraitis}, {Valori}, \&
  {Leake}}]{2018ApJ...865...52L}
{Linan}, L., {Pariat}, {\'E}., {Moraitis}, K., {Valori}, G., \& {Leake}, J.
  2018, \apj, 865, 52

\bibitem[{{Liu} {et~al.}(2018){Liu}, {Cheng}, {Wang}, {Zhou}, {Guo}, \&
  {Cui}}]{2018ApJ...867L...5L}
{Liu}, L., {Cheng}, X., {Wang}, Y., {et~al.} 2018, \apjl, 867, L5

\bibitem[{{Low}(1996)}]{1996SoPh..167..217L}
{Low}, B.~C. 1996, \solphys, 167, 217

\bibitem[{{Metcalf} {et~al.}(2008){Metcalf}, {De Rosa}, {Schrijver}, {Barnes},
  {van Ballegooijen}, {Wiegelmann}, {Wheatland}, {Valori}, \&
  {McTtiernan}}]{2008SoPh..247..269M}
{Metcalf}, T.~R., {De Rosa}, M.~L., {Schrijver}, C.~J., {et~al.} 2008,
  \solphys, 247, 269

\bibitem[{{Moffatt}(1969)}]{1969JFM....35..117M}
{Moffatt}, H.~K. 1969, Journal of Fluid Mechanics, 35, 117

\bibitem[{{Moraitis} {et~al.}(2019){Moraitis}, {Sun}, {Pariat}, \&
  {Linan}}]{2019A&A...628A..50M}
{Moraitis}, K., {Sun}, X., {Pariat}, {\'E}., \& {Linan}, L. 2019, \aap, 628,
  A50

\bibitem[{{Moraitis} {et~al.}(2014){Moraitis}, {Tziotziou}, {Georgoulis}, \&
  {Archontis}}]{2014SoPh..289.4453M}
{Moraitis}, K., {Tziotziou}, K., {Georgoulis}, M.~K., \& {Archontis}, V. 2014,
  \solphys, 289, 4453

\bibitem[{{Pariat} {et~al.}(2017){Pariat}, {Leake}, {Valori}, {Linton},
  {Zuccarello}, \& {Dalmasse}}]{2017A&A...601A.125P}
{Pariat}, E., {Leake}, J.~E., {Valori}, G., {et~al.} 2017, \aap, 601, A125

\bibitem[{{Pariat} {et~al.}(2015){Pariat}, {Valori}, {D{\'e}moulin}, \&
  {Dalmasse}}]{2015A&A...580A.128P}
{Pariat}, E., {Valori}, G., {D{\'e}moulin}, P., \& {Dalmasse}, K. 2015, \aap,
  580, A128

\bibitem[{{Prior} \& {Yeates}(2014)}]{2014ApJ...787..100P}
{Prior}, C. \& {Yeates}, A.~R. 2014, \apj, 787, 100

\bibitem[{{R{\'e}gnier}(2009)}]{2009A&A...497L..17R}
{R{\'e}gnier}, S. 2009, \aap, 497, L17

\bibitem[{{Rust}(1994)}]{1994GeoRL..21..241R}
{Rust}, D.~M. 1994, \grl, 21, 241

\bibitem[{{Scherrer} {et~al.}(2012){Scherrer}, {Schou}, {Bush}, {Kosovichev},
  {Bogart}, {Hoeksema}, {Liu}, {Duvall}, {Zhao}, {Title}, {Schrijver},
  {Tarbell}, \& {Tomczyk}}]{2012SoPh..275..207S}
{Scherrer}, P.~H., {Schou}, J., {Bush}, R.~I., {et~al.} 2012, \solphys, 275,
  207

\bibitem[{{Schrijver} {et~al.}(2006){Schrijver}, {De Rosa}, {Metcalf}, {Liu},
  {McTiernan}, {R{\'e}gnier}, {Valori}, {Wheatland}, \&
  {Wiegelmann}}]{2006SoPh..235..161S}
{Schrijver}, C.~J., {De Rosa}, M.~L., {Metcalf}, T.~R., {et~al.} 2006,
  \solphys, 235, 161

\bibitem[{{Sun}(2013)}]{2013arXiv1309.2392S}
{Sun}, X. 2013, arXiv e-prints, arXiv:1309.2392

\bibitem[{{Taylor}(1974)}]{1974PhRvL..33.1139T}
{Taylor}, J.~B. 1974, \prl, 33, 1139

\bibitem[{{Thalmann} {et~al.}(2011){Thalmann}, {Inhester}, \&
  {Wiegelmann}}]{2011SoPh..272..243T}
{Thalmann}, J.~K., {Inhester}, B., \& {Wiegelmann}, T. 2011, \solphys, 272, 243

\bibitem[{{Thalmann} {et~al.}(2019{\natexlab{a}}){Thalmann}, {Linan}, {Pariat},
  \& {Valori}}]{2019ApJ...880L...6T}
{Thalmann}, J.~K., {Linan}, L., {Pariat}, E., \& {Valori}, G.
  2019{\natexlab{a}}, \apjl, 880, L6

\bibitem[{{Thalmann} {et~al.}(2019{\natexlab{b}}){Thalmann}, {Moraitis},
  {Linan}, {Pariat}, {Valori}, \& {Dalmasse}}]{2019ApJ...887...64T}
{Thalmann}, J.~K., {Moraitis}, K., {Linan}, L., {et~al.} 2019{\natexlab{b}},
  \apj, 887, 64

\bibitem[{{Valori} {et~al.}(2012){Valori}, {D{\'e}moulin}, \&
  {Pariat}}]{2012SoPh..278..347V}
{Valori}, G., {D{\'e}moulin}, P., \& {Pariat}, E. 2012, \solphys, 278, 347

\bibitem[{{Valori} {et~al.}(2013){Valori}, {D{\'e}moulin}, {Pariat}, \&
  {Masson}}]{2013A&A...553A..38V}
{Valori}, G., {D{\'e}moulin}, P., {Pariat}, E., \& {Masson}, S. 2013, \aap,
  553, A38

\bibitem[{{Valori} {et~al.}(2016){Valori}, {Pariat}, {Anfinogentov}, {Chen},
  {Georgoulis}, {Guo}, {Liu}, {Moraitis}, {Thalmann}, \&
  {Yang}}]{2016SSRv..201..147V}
{Valori}, G., {Pariat}, E., {Anfinogentov}, S., {et~al.} 2016, \ssr, 201, 147

\bibitem[{{Wheatland} {et~al.}(2000){Wheatland}, {Sturrock}, \&
  {Roumeliotis}}]{2000ApJ...540.1150W}
{Wheatland}, M.~S., {Sturrock}, P.~A., \& {Roumeliotis}, G. 2000, \apj, 540,
  1150

\bibitem[{{Wiegelmann}(2004)}]{2004SoPh..219...87W}
{Wiegelmann}, T. 2004, \solphys, 219, 87

\bibitem[{{Wiegelmann}(2008)}]{2008JGRA..113.3S02W}
{Wiegelmann}, T. 2008, Journal of Geophysical Research (Space Physics), 113,
  A03S02

\bibitem[{{Wiegelmann} \& {Inhester}(2010)}]{2010A&A...516A.107W}
{Wiegelmann}, T. \& {Inhester}, B. 2010, \aap, 516, A107

\bibitem[{{Wiegelmann} {et~al.}(2006){Wiegelmann}, {Inhester}, \&
  {Sakurai}}]{2006SoPh..233..215W}
{Wiegelmann}, T., {Inhester}, B., \& {Sakurai}, T. 2006, \solphys, 233, 215

\bibitem[{{Wiegelmann} \& {Sakurai}(2012)}]{2012LRSP....9....5W}
{Wiegelmann}, T. \& {Sakurai}, T. 2012, Living Reviews in Solar Physics, 9, 5

\bibitem[{{Wiegelmann} {et~al.}(2012){Wiegelmann}, {Thalmann}, {Inhester},
  {Tadesse}, {Sun}, \& {Hoeksema}}]{2012SoPh..281...37W}
{Wiegelmann}, T., {Thalmann}, J.~K., {Inhester}, B., {et~al.} 2012, \solphys,
  281, 37

\bibitem[{{Wiegelmann} {et~al.}(2008){Wiegelmann}, {Thalmann}, {Schrijver}, {De
  Rosa}, \& {Metcalf}}]{2008SoPh..247..249W}
{Wiegelmann}, T., {Thalmann}, J.~K., {Schrijver}, C.~J., {De Rosa}, M.~L., \&
  {Metcalf}, T.~R. 2008, \solphys, 247, 249

\bibitem[{{Woltjer}(1958)}]{1958PNAS...44..833W}
{Woltjer}, L. 1958, Proceedings of the National Academy of Science, 44, 833

\bibitem[{{Zou} {et~al.}(2020){Zou}, {Jiang}, {Wei}, {Feng}, {Zuo}, \&
  {Wang}}]{2020ApJ...890...10Z}
{Zou}, P., {Jiang}, C., {Wei}, F., {et~al.} 2020, \apj, 890, 10

\bibitem[{{Zuccarello} {et~al.}(2018){Zuccarello}, {Pariat}, {Valori}, \&
  {Linan}}]{2018ApJ...863...41Z}
{Zuccarello}, F.~P., {Pariat}, E., {Valori}, G., \& {Linan}, L. 2018, \apj,
  863, 41

\end{thebibliography}

\end{document}